\documentclass[11pt,letterpaper]{article}
\pdfoutput=1
\usepackage{jheppub}
\usepackage{color}
\usepackage{graphicx}
\usepackage{wrapfig}

\usepackage{verbatim}
\usepackage{amsmath}
\usepackage{amssymb}
\usepackage{subfig}
\usepackage{url}
\usepackage{bbold}
\usepackage{xspace}

\usepackage{multirow}
\usepackage{threeparttable}
\usepackage{paralist}

\usepackage{color}

\usepackage{floatrow}
\newfloatcommand{capbtabbox}{table}[][\FBwidth]

\usepackage{tikz}
\usetikzlibrary{trees}
\usetikzlibrary{decorations.pathmorphing}
\usetikzlibrary{decorations.markings}
\usetikzlibrary{arrows}

\definecolor{darkgreen}{rgb}{0,0.5,0}

\newcommand{\CO}{\mathcal{O}}
\newcommand{\CZ}{\mathcal{Z}}
\newcommand{\Dfbd}{\mathord{\buildrel{\lower3pt\hbox{$\scriptscriptstyle\leftrightarrow$}}\over {D}_{\mu}}}

\newcommand{\beq}{\begin{equation}}
\newcommand{\eeq}[1]{\label{#1}\end{equation}}
\def\beqa{\begin{eqnarray}}
\def\eeqa#1{\label{#1}\end{eqnarray}}
\newcommand{\eeqn}{\end{equation}}
\newcommand{\CR}{\notag \\}
\newcommand{\leqn}[1]{(\ref{#1})}

\newcommand{\HZ}{e^+e^-\rightarrow hZ}
\newcommand{\cW}{c_\theta}
\newcommand{\sW}{s_\theta}

\def\stacksymbols #1#2#3#4{\def\theguybelow{#2}
    \def\vp{\lower#3pt}
    \def\sp{\baselineskip0pt\lineskip#4pt}
    \mathrel{\mathpalette\intermediary#1}}

\def\intermediary#1#2{\vp\vbox{\sp
     \everycr={}\tabskip0pt
     \halign{$\mathsurround0pt#1\hfil##\hfil$\crcr#2\crcr
              \theguybelow\crcr}}}

\def\lsim{\stacksymbols{<}{\sim}{2.5}{.2}}
\begin{document}

\tikzset{
	  photon/.style={decorate, decoration={snake}, draw=black},
	  boson/.style={decorate, decoration={snake}, draw=black},
	  electron/.style={draw=black, postaction={decorate},
	           decoration={markings,mark=at position .55 with {\arrow[draw=black]{latex}}}
	  },
	  electron2/.style={draw=black, postaction={decorate},
	           decoration={markings,mark=at position .55 with {\arrow[draw=black]{latex reversed}}}
	  },
	  fermion/.style={draw=black, postaction={decorate},
	            decoration={markings,mark=at position .55 with {\arrow[draw=black]{}}}
	  },
 pttt/.style={decorate, draw=white},
	  gluon/.style={decorate, draw=black, 
	    decoration={coil,amplitude=4pt, segment length=6pt}},
	  gluon/.style={decorate, draw=black, 
	    decoration={coil,amplitude=4pt, segment length=6pt}},
	  higgs/.style={draw=black, postaction={decorate},
	           decoration={markings,mark=at position .55 with}
	  },
	  nothing/.style={draw=white}
	}

\title{Precision Higgsstrahlung as a Probe of New Physics}

\author[a]{Nathaniel Craig,}
\author[b]{Marco Farina,}
\author[c]{Matthew McCullough,}
\author[b]{and Maxim Perelstein}

\affiliation[a]{Department of Physics, University of California, Santa Barbara, CA 93106, USA}
\affiliation[b]{Laboratory for Elementary Particle Physics, Cornell University, Ithaca, NY 14853, USA}
\affiliation[c]{Theory Division, CERN, 1211 Geneva 23, Switzerland}

\emailAdd{ncraig@physics.ucsb.edu}
\emailAdd{mf627@cornell.edu}
\emailAdd{matthew.mccullough@cern.ch}
\emailAdd{mp325@cornell.edu}

\date{\today}

\abstract{A ``Higgs factory", an electron-positron collider with center-of-mass energy of about 250 GeV, will measure the cross section of the Higgsstrahlung process, $\HZ$, with sub-percent precision. This measurement is sensitive to a variety of new physics scenarios. In this paper, we study two examples. First, we compute corrections to the $\HZ$ differential cross section in the effective field theory (EFT) approach, including the complete set of dimension-6 operators contributing to this process. These results are applicable to any model where the new physics mass scale is significantly above the weak scale. Second, we present a complete one-loop calculation of the effect of third-generation squarks, with arbitrary soft masses and mixing, on this cross section. This is expected to be the leading correction in natural supersymmetric models. We demonstrate the agreement between the full one-loop calculation and the EFT result in the limit of large stop masses. Finally, we estimate the discovery reach of the $\HZ$ cross section measurement in the two models.}



\preprint{}

\maketitle

\section{Introduction}

Experimental determination of the Higgs boson properties will be the major focus of high-energy physics in the coming years. The well-known naturalness argument suggests that the Standard Model (SM) picture of electroweak symmetry breaking (EWSB) is incomplete, and new physics in the electroweak sector has to appear at energy scale of $\sim$TeV or less. Given that the Higgs boson plays a central role in EWSB, it is natural to expect that this new physics will influence its properties, leading to deviations from the SM predictions. A program of precision measurements of Higgs properties offers an exciting opportunity to search for such effects.

Some of the Higgs properties are already being constrained by the LHC experiments. For example, the Higgs couplings to $W/Z$ bosons and gluons have been measured and found to agree with the SM to within about 20\%~\cite{CMS-PAS-HIG-14-009}. The impending Run-2 of the LHC, and the future luminosity upgrade program, will both improve the precision of these measurements and measure additional couplings. However, the composite nature of the proton makes it difficult to reduce the systematic and theory errors below a few \%. In addition, interpretation of the LHC rate measurements in terms of couplings is somewhat model-dependent, as it depends on the total width of the Higgs which is not directly observable. To go to the next level of precision, an electron-positron
collider will be required. Currently, plans for constructing such a ``Higgs factory" are under serious consideration~\cite{Baer:2013cma,Gomez-Ceballos:2013zzn,Dawson:2013bba}. The Higgs factory would run at a center-of-mass energy of about 250 GeV, where Higgs boson production is dominated by the Higgsstrahlung process, $\HZ$. A multi-year experimental program is envisioned, in which a combined data set of $\CO(10^5-10^6)$ Higgsstrahlung events would be accumulated. Experimentally, these events are very clean. They can be tagged by the $Z$ boson energy, without necessarily reconstructing the Higgs, providing a robust, model-independent cross section measurement. Moreover, very precise SM predictions can be obtained for this purely electroweak process. All these factors combine to make the $\HZ$ cross section a uniquely powerful observable in a search for physics beyond the SM. In this paper, we investigate the potential sensitivity of this measurement to new physics, assuming that precision of $\CO(0.1-0.5\%)$ can ultimately be reached at a Higgs factory.

The paper is organized as follows: In Section~\ref{sec:sm} we review the Higgsstrahlung process in the Standard Model and discuss the appropriate observables for characterizing deviations from SM predictions for $\HZ$ in the presence of new physics. In Section~\ref{sec:eft} we frame new physics contributions to Higgsstrahlung in the language of Effective Field Theory (EFT), enumerating a complete basis of CP-conserving dimension-6 operators relevant to the $\HZ$ process and computing their respective contributions to the $\HZ$ cross section.
The effect of anomalous Higgs couplings on this process has been considered before~\cite{Hagiwara:1993sw,Gounaris:1995mx,Kilian:1996wu,GonzalezGarcia:1999fq,Hagiwara:2000tk,Barger:2003rs,Biswal:2005fh,Kile:2007ts,Dutta:2008bh,Contino:2013gna,Amar:2014fpa}. However, most of these studies did not include the complete set of operators; typically, the operators already constrained by precision electroweak measurements were omitted. In addition, none of these studies included the effect of the shift in the couplings entering the leading-order SM prediction for the $\HZ$ cross section, relative to their reference values, due to the effect of dimension-6 operators on the electroweak input observables ({\it e.g.} the $Z$ mass, $G_F$, and $\alpha$). Again, this was justified by the tight constraints on such shifts from precision electroweak fits; also, many studies focused explicitly on angular distributions, which, unlike the total rate, are unaffected by such coupling shifts. Since the currently discussed Higgs factories would be able to measure the Higgsstrahlung cross section with precision approaching that of the best precision electroweak observables, these effects need to be taken into account to properly interpret this measurement. We will do so in this paper. (Note that another study of the $\HZ$ process with the complete operator set and proper inclusion of the SM coupling shifts has recently appeared; see Ref.~\cite{Beneke:2014sba}.)
We also estimate the energy scales that can be probed by the Higgs factory Higgsstrahlung measurement interpreted in the EFT approach. In Section \ref{sec:NLO} we consider a specific realization of new physics in the form of third-generation squarks in supersymmetric models. We perform both a full next-to-leading order (NLO) calculation\footnote{For earlier work considering the NLO corrections to the Higgsstrahlung process in the presence of new physics, including supersymmetric theories, see Refs.~\cite{Chankowski:1992er,Driesen:1995ew,Heinemeyer:2001iy,Dawson:2002wc,Englert:2013tya,Gounaris:2014tha,Cao:2014rma}.} of $\HZ$ and the corresponding EFT calculation of $\HZ$ in the presence of third-generation squarks. To this end we make use of the recent calculation of Wilson coefficients for third generation squarks~\cite{Henning:2014gca}. The excellent agreement between our NLO and EFT results in the limit of large squark mass serves as a highly non-trivial check of both calculations, while the full NLO result allows us to establish the range of validity of the EFT approach as the squark mass is lowered. We also discuss the reach of the Higgs factory indirect searches for stops, including both Higgsstrahlung and the measurement of $h\to gg, \gamma\gamma$ decay widths. Finally, we present our conclusions in Section~\ref{sec:conc}.

\section{$\HZ$ in the Standard Model}
\label{sec:sm}

To define notation and set the stage for subsequent discussion, let us briefly review the well-known results for the Higgsstrahlung process in the Standard Model (SM).
The differential cross section is given by
\beq
\frac{d\sigma_{\rm SM}}{d\cos\theta} = \frac{p_Z}{16\pi s^{3/2}} \,F_{\rm SM}(s, t),
\eeq{xsecSM}
where $\theta$ is the angle between the electron beam and the $Z$ momentum and
\beq
p_Z = \frac{\sqrt{s}}{2}\,\left( 1-\frac{(m_h+m_Z)^2}{s}\right)^{1/2} \left( 1-\frac{(m_h-m_Z)^2}{s}\right)^{1/2}
\eeq{pZ}
is the $Z$ boson momentum in the center-of-mass frame of the collision. Assuming unpolarized beams, a tree level calculation yields~\cite{Ellis:1975ap,Ioffe:1976sd,Bjorken:1977wg,Kilian:1995tr}
\beq
F_{\rm SM}(s, t) = \frac{1}{4}g_{\rm \tiny ZZh}^2 (g_L^2+g_R^2) \frac{2s + \frac{tu}{m_Z^2}-m_h^2}{(s-m_Z^2)^2}\,,
\eeq{Msq}
where $g_L$ and $g_R$ are the couplings of the left-handed and right-handed electrons to the $Z$ boson.
Here we used the standard Mandelstam variables:
\beqa
s &=& (p_1+p_2)^2,~~~
t = (p_1-p_4)^2 = m_Z^2-\sqrt{s}(E_Z-p_Z\cos\theta), \CR
u &=& (p_1-p_3)^2=m_Z^2-\sqrt{s}(E_Z+p_Z\cos\theta) = m_Z^2+m_h^2-s-t,
\eeqa{Mandelstam}
where
\beq
E_Z = \frac{s+m_Z^2-m_h^2}{2\sqrt{s}}
\eeq{EzPz}
is the $Z$ energy in the c-o-m frame. For reference, the numerical value of the tree-level SM cross section at $\sqrt{s}=250$ GeV is 224 fb.

In the context of our study, the coupling constants appearing in Eq.~\leqn{Msq} deserve a careful discussion. Potential precision of the $\HZ$ cross section measurement at the Higgs factories, of order $0.1$\%, matches or surpasses that achieved in precision electroweak (PEW) experiments at the $Z$ pole. A comparison of the SM with experiment at this level requires that all numerical inputs into the SM prediction be known to at least the same precision. The standard approach, well-known in the case of PEW analyses, is to use three most precisely measured electroweak-sector observables as inputs, infer the ``reference" values of the SM Lagrangian parameters from these inputs, and compute the numerical values of all other observables using these reference values. We will adopt the same approach. Specifically, we will consider two sets of inputs, or ``bases". In the first basis, we take the $Z$ mass, the fine-structure constant $\alpha$ at zero momentum transfer, and the Fermi constant $G_F$ inferred from muon decay rate, as inputs. In the second basis, we use the $W$ mass instead of the Fermi constant.\footnote{While $G_F$ has been measured to a much higher precision than $m_W$, the second basis will be important for a comparison of the full one-loop and effective field theory calculations in Section~\ref{sec:compare}.} The reference values of the relevant couplings are given by
\beq
\hat{g}_{\rm \tiny ZZh} = \hat{g}_z m_Z,~~\hat{g}_L = \hat{g}_z\left( -\frac{1}{2}+\sin^2\hat{\theta}_W \right),~~\hat{g}_R =  \hat{g}_z \sin^2\hat{\theta}_W,
\eeq{gvalues}
where
\beq
\hat{g}_z = \frac{4\sqrt{\pi\alpha}}{\sin2\hat{\theta}_W}
\eeq{gz}
and the reference value of the Weinberg angle depends on the basis:
\beqa
(m_Z, G_F, \alpha):&~& \sin2\hat{\theta}_W = \left( \frac{4\pi\alpha}{\sqrt{2}G_F m_Z^2}\right)^{1/2},\CR
(m_Z, m_W, \alpha):&~&\cos\hat{\theta}_W=\frac{m_W}{m_Z}\,.
\eeqa{g_refs}
If the SM is the true theory, the numerical value of the SM cross section obtained with these two input bases, or indeed any other basis, are identical within the experimental errors on the inputs. However, if there is new physics, it may affect the observables used to define the reference couplings. In this case, the true values of the couplings $g_i$ in the SM Lagrangian differ from their reference values $\hat{g}_i$:
\beq
g_{\rm \tiny ZZh} =\hat{g}_{\rm \tiny ZZh} + \delta g_{\rm \tiny ZZh},~~g_L = \hat{g}_L + \delta g_L,~~g_R = \hat{g}_R + \delta g_R.
\eeq{delta_gs}
To search for new physics in $\HZ$, one would compare the experimentally measured cross section $\sigma_{\rm exp}$ with the reference SM cross section $\sigma_{\rm SM}(\hat{g}_i)$. The apparent cross section shift
\beq
\Delta\sigma \equiv \sigma_{\rm exp} - \sigma_{\rm SM}(\hat{g}_i)
\eeq{xsshift}
should thus incorporate the effect of coupling shifts $\delta g_i$, as well as the direct contribution of new physics to the $\HZ$ cross section. In the presence of new physics, the reference SM cross section values obtained in different bases are no longer the same. Since $\sigma_{\rm exp}$ is physically observable and therefore must be basis-independent, this leads to basis dependence of the cross section shift $\delta\sigma$. We will observe this dependence in our explicit calculations of $\delta\sigma$ in the following section. It should be emphasized that bounds on new physics obtained in different bases must be identical, as long as a global fit to all available observables is performed in each case and the uncertainties in the input observables are properly taken into account.

Next-to-Leading-Order (NLO) corrections to the Higgsstrahlung cross section in the SM are well-known \cite{Fleischer:1982af,Kniehl:1991hk,Denner:1992bc}.  For a Higgs mass of $125$ GeV and CM energy $\sqrt{s}=250$ GeV the full NLO electroweak corrections amount to a $3\%$ shift in the Higgstrahlung cross section relative to the LO result.  While small, such corrections are within the realm of proposed future colliders.  NNLO electroweak and mixed QCD-electroweak corrections have not yet been calculated, although they are likely to constitute the dominant source of theoretical uncertainty.  In this paper, we will assume that a sufficiently precise SM prediction will be available to bring the theory uncertainty to a level subdominant to the statistical error in the cross section measurement.

\section{Effective Field Theory Approach}
\label{sec:eft}

The effects of any new physics appearing at a mass scale $\Lambda$ on the Higgsstrahlung cross section can be described in terms of an Effective Field Theory (EFT), as long as $\Lambda$ is large compared to the center-of-mass energy $\sqrt{s}$ and the weak scale $v$. In general, the EFT Lagrangian is an expansion in inverse powers of $\Lambda$. The term of order $\Lambda^{-n}$ contains all possible operators of mass dimension $4+n$ compatible with the symmetries imposed on the theory, in our case the full SM gauge symmetry as well as lepton and baryon number. With these restrictions, the leading term in the expansion is $n=2$, containing dimension-6 operators.

A complete set of CP-conserving dimension-6 operators that can contribute to the $\HZ$ process is listed in Table~\ref{tab:hit}. This basis is complete in the sense that an arbitrary set of CP-conserving dimension-6 operators contributing to the $\HZ$ process can be reduced to the operators listed in Table~\ref{tab:hit} (plus additional operators irrelevant to $\HZ$) by field redefinitions. In principle there can be additional contributions from the dipole-type operators  $\CO_{DB}^e \sim \bar L_L \sigma^{\mu \nu} e_R H B_{\mu \nu}$, $\CO_{DW}^e \sim \bar L_L \sigma^{\mu \nu} e_R \sigma^a H W_{\mu \nu}^a$. However, these are expected to be Yukawa-suppressed due to the chirality flip; moreover, since they do not interfere with the SM amplitude, their leading contribution is of order $1/\Lambda^4$. We therefore do not include them in this analysis. This completes the enumeration of CP-conserving dimension-6 operators contributing to $\HZ$.

The dimension-6 Lagrangian has the form
\beq
{\cal L}_{\rm \tiny pre-EWSB} = \sum_i \frac{c_i}{\Lambda^2} \, \CO_i\,,
\eeq{LpreEW}
where $c_i$ are dimensionless Wilson coefficients. Given a complete theory at the scale $\Lambda$, the Wilson coefficients can be computed in terms of the parameters of that theory; we will consider an example of this in Section~\ref{sec:compare}. In this section, we treat $c_i$'s as free parameters.

\begin{table}[t]
\begin{center}
\begin{tabular}{|c|}
\hline
$\CO_{WW} = g^2 |H|^2 W_{\mu \nu}^a W^{a,\mu \nu}$   \\
$\CO_{BB} = g'^2 |H|^2 B_{\mu \nu} B^{\mu \nu}$  \\
$\CO_{WB} = g g' H^\dag \sigma^a H W_{\mu \nu}^a B^{\mu \nu}$   \\
$\CO_H = \frac{1}{2} (\partial_\mu |H|^2)^2$ \\
$\CO_T = \frac{1}{2}(H^\dag \Dfbd H)^2$ \\
$\CO_L^{(3)\ell} = (i H^\dag \sigma^a \Dfbd H) (\bar L_L \gamma^\mu \sigma^a L_L)$ \\
$\CO_{LL}^{(3)\ell} = (\bar L_L  \gamma_\mu \sigma^a L_L) (\bar L_L \gamma^\mu \sigma^a L_L)$ \\
$\CO_L^\ell = (i H^\dag \Dfbd H)(\bar L_L \gamma^\mu L_L)$ \\
$\CO_R^e = (i H^\dag \Dfbd H)(\bar e_R \gamma^\mu e_R)$ \\
\hline
\end{tabular}
\end{center}
\caption{A complete set of CP-conserving dimension-6 operators which contribute to $\HZ$.}
\label{tab:hit}
\end{table}%

To account for electroweak symmetry breaking, we write the Higgs doublet as $H=\left(0, \frac{v+h}{\sqrt{2}}\right)^T$, where $v\approx 246$ GeV is the Higgs vev, while $h$ is the physical Higgs boson field. Note that in the presence of new physics, the field $h$ defined in this way is not canonically normalized; to return to canonical normalization requires a field redefinition
\beq
h \rightarrow \left( 1 + \frac{c_H}{2 \Lambda^2} v^2\right) h.
\eeq{redef}
Performing this field redefinition and dropping the terms that do not contribute to $\HZ$, the dim.-6 Lagrangian of Eq.~\leqn{LpreEW} reduces to 
\beqa
{\cal L}_{\rm \tiny post-EWSB} &=& \frac{d_1 v^3}{2\Lambda^2} h Z_\mu Z^\mu + \frac{d_2 v}{4\Lambda^2} h \CZ_{\mu\nu} \CZ^{\mu\nu} + 
\frac{d_3 v}{2\Lambda^2} h F^{\mu\nu}\CZ_{\mu\nu} 
\CR & & + \frac{v}{\Lambda^2} \bar{\psi} \gamma^\mu \left( d_4 P_L + d_5 P_R \right) \psi Z_\mu h,
\eeqa{LpostEW}
where $Z_\mu$ is the SM $Z$ boson field, $\CZ_{\mu\nu}=\partial_\mu Z_\nu- \partial_\nu Z_\mu$, $F_{\mu\nu}$ is the electromagnetic field strength tensor, $\psi$ is the (Dirac) electron field, and $P_R=\frac{1}{2}(1+\gamma_5)$ and $P_L=\frac{1}{2}(1-\gamma_5)$ are helicity projectors. The Feynman rules derived from this Lagrangian are

\begin{tabular}{l l}
  \parbox{0.2\textwidth}{
  \begin{tikzpicture}[
        thick,
        level/.style={level distance=1.2cm},
        level 2/.style={sibling distance=2cm},
        level 3/.style={sibling distance=1.2cm}
    ]
    \coordinate
        child[grow=left]{
            child {
                node {$Z_{\mu}$}
                edge from parent [photon, solid]
                node[above, pos=0.35] {$\searrow$}
                node[above right, pos=0.67] {$k_1$}
            }
            child {
                node {$Z_{\nu}$}
                edge from parent [photon, solid]
                      node[below, pos=0.35] {$\nearrow$}
                node[below right, pos=0.67] {$k_2$}
            }
            edge from parent [higgs, dashed]
            node [pos=-.15] {$h$}
            node[above] {$\leftarrow$}
            node[above right, pos=0.45] {$k_h$}
        }

    ;
\end{tikzpicture}
  }
   &
   \parbox{0.71\textwidth}{
\vspace{1.1cm}
\beq
=i g_{\rm \tiny ZZh}g^{\mu\nu}+  \frac{iv}{\Lambda^2} \Bigl[ g^{\mu\nu} \left(v^2 d_1 -(k_1\cdot k_2) d_2  \right)+ k_1^\nu k_2^\mu d_2 
\eeq{Frule}
}
 \\
 \parbox{0.2\textwidth}{

\begin{tikzpicture}[
        thick,
        level/.style={level distance=1.2cm},
        level 2/.style={sibling distance=2cm},
        level 3/.style={sibling distance=1.2cm}
    ]
    \coordinate
        child[grow=left]{
            child {
                node {$Z_{\mu}$}
                edge from parent [photon, solid]
                node[above, pos=0.35] {$\searrow$}
                node[above right, pos=0.67] {$k_1$}
            }
            child {
                node {$\gamma_{\nu}$}
                edge from parent [photon, solid]
                      node[below, pos=0.35] {$\nearrow$}
                node[below right, pos=0.67] {$k_2$}
            }
            edge from parent [higgs, dashed]
            node [pos=-.15] {$h$}
            node[above] {$\leftarrow$}
            node[above right, pos=0.45] {$k_h$}
        }

    ;
\end{tikzpicture}

    }
  &

\parbox{0.71\textwidth}{
\beq
=\frac{iv d_3}{\Lambda^2}  \Bigl[  -(k_1\cdot k_2) g^{\mu\nu}   + k_1^\nu k_2^\mu  \Bigr] 
\eeq{Frule2}
}
\\
  \parbox{0.2\textwidth}{

\begin{tikzpicture}[
        thick,
        level/.style={level distance=1.2cm},
        level 2/.style={sibling distance=2cm},
        level 3/.style={sibling distance=1.2cm}
    ]
    \coordinate
        child[grow=left]{
            child {
                node {$e^+$}
                edge from parent [electron, solid]
            }
            child {
                node {$e^-$}
                edge from parent [electron2,solid]
            }
            edge from parent [pttt]
        }
        child[grow=right, level distance=-1.2cm] {
        child  {
        node  {$h$}
            edge from parent [higgs,dashed]
        }
        child {
        node {$Z_\mu$}
            edge from parent [photon, solid]
        }
                            edge from parent [pttt]
    };
\end{tikzpicture}

    }
  &

\parbox{0.71\textwidth}{
\beq
=\frac{iv}{\Lambda^2} \gamma^\mu \left( d_4 P_L + d_5 P_R \right)~~~~~~~~~~~~~~~~~~~~~~~~~~~~~~~~~~~~~~~~~~~~~~~~~~~~
\eeq{Frule2}
}
\end{tabular}

\vskip.5cm

The dimensionless coefficients $d_i$ are given by
\beqa
d_1 &=& - \frac{g_z^2}{2} \left( \frac{1}{2}c_H + 2 c_T\right), \CR
d_2 &=& 4g_z^2 \left(\sW^4 c_{BB} + \sW^2 \cW^2 c_{WB} + \cW^4 c_{WW}\right), \CR
d_3 &=& 2g_z^2 \cW \sW\left(-2\sW^2 c_{BB} - (\cW^2-\sW^2) c_{WB} + 2 \cW^2 c_{WW}\right), \CR
d_4 &=& -g_z \left( c_L^{(3)\ell} + c_L^\ell \right),\CR
d_5 &=& -g_z c^e_R,
\eeqa{deltas}
where we used the shorthand notation $\cW\equiv \cos\theta_W$, $\sW\equiv\sin\theta_W$, and the coupling $g_z$ defined in Eq.~\leqn{gz}.\footnote{Note that the difference between the actual and reference values of $g_z$, discussed in the previous section, is irrelevant in these formulas. This difference amounts to corrections of order $\Lambda^{-4}$ to physical observables, {\it i.e.}~of the same order as dim.-8 operators that we ignored.} Our expressions for the $d_i$ are in excellent agreement with EFT results for comparable bases in the existing literature (e.g. \cite{Giudice:2007fh, Contino:2013kra, Elias-Miro:2013mua, Pomarol:2013zra, Elias-Miro:2013eta}). 

In addition, upon electroweak symmetry breaking, the dim.-6 operators in~\leqn{LpreEW} induce shifts between the coupling constants in the SM Lagrangian and their reference values, as explained in Section~\ref{sec:sm}. In the two bases of interest, we obtain:\footnote{We are grateful to Michael Fedderke for pointing out an error in the first set of formulas in Eq.~\leqn{shiftsMW} in the original version of this paper.} 
\beqa
(m_Z, G_F, \alpha):
&~& \delta g_{\rm \tiny ZZh}=  \frac{g_{\rm \tiny ZZh} v^2}{\Lambda^2} \left( c_T - c_L^{(3)\ell} + c_{LL}^{(3)\ell}\right),\CR
&~&  \delta g_L =  \frac{g_z v^2}{2\Lambda^2} \left[ -\frac{1}{2(\cW^2-\sW^2)} c_T + \frac{2e^2}{\cW^2-\sW^2}c_{WB} +\frac{2\sW^2}{\cW^2-\sW^2} c_L^{(3)\ell} -\frac{1}{\cW^2-\sW^2}c_{LL}^{(3)\ell} -c_L^\ell \right],   \CR
&~&  \delta g_R = \frac{g_z v^2}{2\Lambda^2} \left[ -\frac{\sW^2}{\cW^2-\sW^2} c_T + \frac{2e^2}{\cW^2-\sW^2}c_{WB} + \frac{2\sW^2}{\cW^2-\sW^2} \left( c_L^{(3)\ell} -c_{LL}^{(3)\ell}\right) -  c^e_R \right]; \CR\CR
(m_Z, m_W, \alpha):
&~& \delta g_{\rm \tiny ZZh}= \frac{g_{\rm \tiny ZZh} v^2}{\Lambda^2} \left[ \left( 1-\frac{\cW^2}{2\sW^2}\right)c_T + g^2 c_{WB} \right],\CR
&~&  \delta g_L = \frac{g_z v^2}{2\Lambda^2} \left[ \frac{1}{2\sW^2}\, c_T - g^2 c_{WB} -c_L^{(3)\ell}-c_L^\ell  \right],\CR
&~&  \delta g_R = \frac{g_z v^2}{2\Lambda^2} \left[ c_T - c^e_R \right].
\eeqa{shiftsMW}
The $\HZ$ cross section shift with respect to the reference SM cross section, as defined in Eq.~\leqn{xsshift}, is given by
\beq
\frac{d\Delta \sigma}{d\cos\theta} = \frac{p_Z}{16\pi s^{3/2}} \,\left[ 2\left( \frac{\delta g_{\rm \tiny ZZh}}{g_{\rm \tiny ZZh}} + \frac{g_L \delta g_L + g_R \delta g_R}{g_L^2+g_R^2}\right) F_{\rm SM} (s, t) + \frac{g_{\rm \tiny ZZh} v}{2\Lambda^2}\sum_{i=1}^5 d_i F_i(s, t) \right],
\eeq{xsec_direct}
where the functions $F_i$ are collected in Table~\ref{tab:Fs}. The first term in the square brackets reflects the effect of the coupling constant shifts, while the second term is the ``direct" contribution of new interactions, Eq.~\leqn{LpostEW}, to this cross section. The direct contribution is due to the interference between the SM diagrams and those with a single $d_i$ insertion.

\begin{table}[t]
\begin{center}
\begin{tabular}{|r|c|}
\hline
$F_1$ & $(g_L^2+g_R^2) v^2 \frac{2s + \frac{tu}{m_Z^2}-m_h^2}{(s-m_Z^2)^2}$   \\
\hline
$F_2$ & $(g_L^2+g_R^2) \frac{s (s+m_Z^2-m_h^2)}{(s-m_Z^2)^2}$ \\
\hline
$F_3$ & $-e(g_L+g_R) \frac{s+m_Z^2-m_h^2}{s-m_Z^2}$ \\
\hline
$F_4$ & $g_L \frac{2s + \frac{tu}{m_Z^2}-m_h^2}{s-m_Z^2}$ \\
\hline
$F_5$ & $g_R \frac{2s + \frac{tu}{m_Z^2}-m_h^2}{s-m_Z^2}$ \\
\hline
\end{tabular}
\end{center}
\caption{Direct contributions to the $\HZ$ differential cross section from each operator in the post-EWSB Lagrangian.}
\label{tab:Fs}
\end{table}%

\begin{figure}[h]
\CenterFloatBoxes
\begin{floatrow}
\capbtabbox
  {\begin{tabular}{|r|c|c|c|c|}
\hline
 & $\delta\sigma/\sigma=0.5$\% & $\delta\sigma/\sigma=0.1$\% & PEW  & LHC\\
\hline
$\CO_{WW}$ & 5.1/3.2 & 11.5/7.5 & - & 2.5\\
$\CO_{BB}$ &  1.0/0.64 & 2.2/1.4&  - & 2.5 \\
$\CO_{WB}$  & 2.1/1.3 & 4.6/2.9 & 0.3 & 2.5\\
$\CO_H$ & 2.5/1.6 & 5.5/3.5 & -& - \\
$\CO_T$ & 2.0/1.3 & 4.5/2.8 & 1.0& - \\
$\CO_L^{(3)\ell}$ & 8.6/5.4 & 19/12 & 1.2 & -\\
$\CO_{LL}^{(3)\ell}$ & 5.3/3.4 & 12/7.5 & 4.3& -\\
$\CO_L^\ell$ & 10.1/6.4 & 23/14 & 1.5 & -\\
$\CO_R^e$ & 8.7/5.5 & 19/12& 1.0 &-\\
\hline
\end{tabular}
  }
    {\caption{Exclusion (95\%~c.l.)/discovery (5-sigma) reach of a measurement of $\sigma(\HZ)$ at $\sqrt{s}=250$ GeV. The reach is in terms of $\Lambda/\sqrt{c_i}$, in TeV, for each operator $\CO_i$. For comparison, current precision electroweak bounds from Ref.~\cite{Falkowski:2014tna} and LHC bounds from $h\gamma\gamma$ effective coupling measurement~\cite{Khachatryan:2014jba} are also shown.}\label{tab:reach}}
\ffigbox
  {\includegraphics[height=0.6\textwidth]{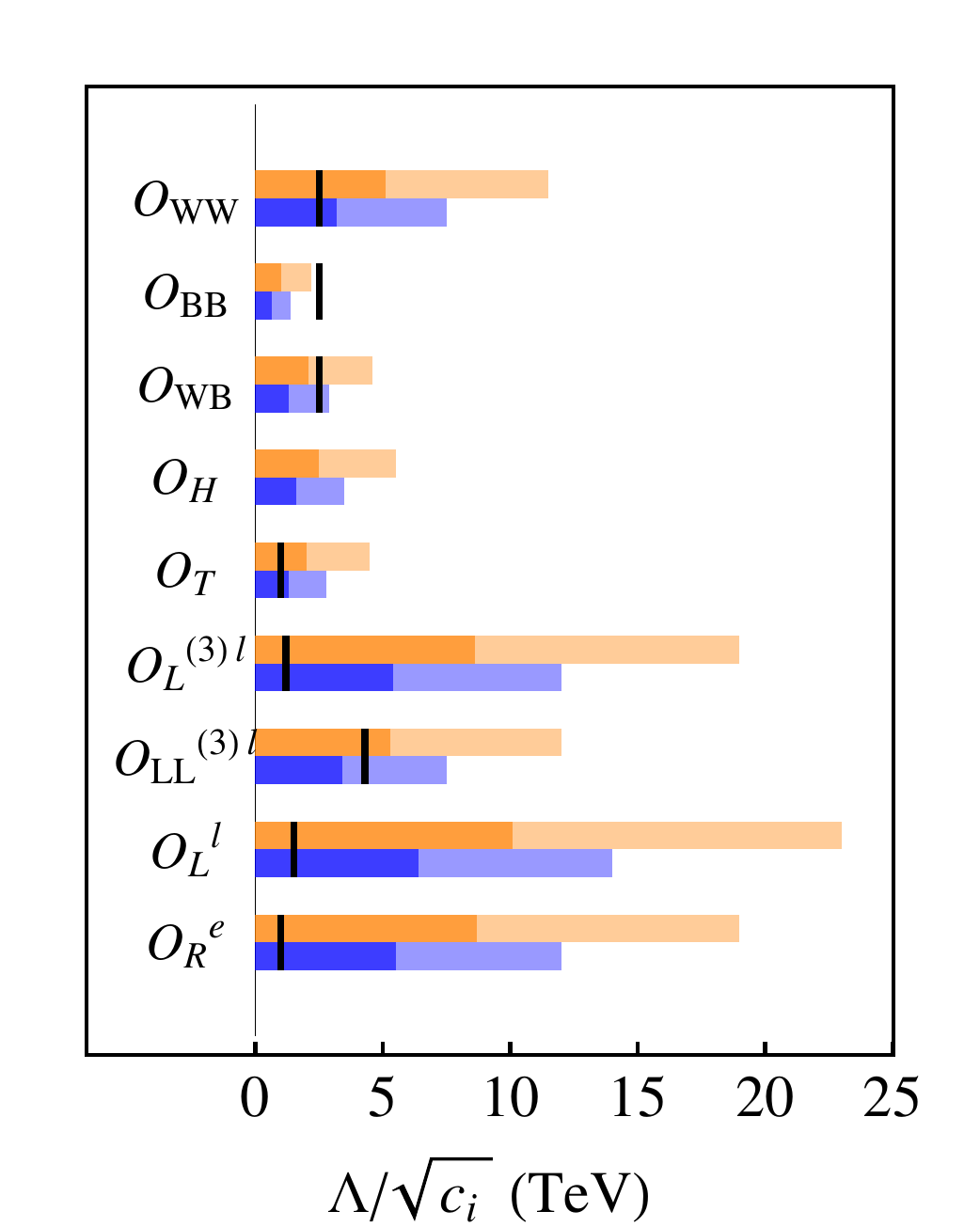}}
  {\caption{Graphical representation of the results in Table~\ref{tab:reach}. The exclusion reach is shown in orange and the discovery reach in blue (paler colors for higher accuracy). Black lines denote the current precision electroweak and LHC bounds.}
  \label{fig:reach}}
\end{floatrow}
\end{figure}

The fractional deviation of the total cross section from its reference SM value, in the $(m_Z, G_F, \alpha)$ basis and at $\sqrt{s}=250$ GeV, is approximately given by 
\beqa
\frac{\Delta\sigma}{\sigma} &\approx& \Bigl( 0.26 c_{WW} + 0.01 c_{BB} +0.04 c_{WB} 
-0.06 c_H -0.04 c_T +0.74 c_L^{(3)\ell} \CR & & + 0.28 c_{LL}^{(3)\ell} + 1.03 c_L^\ell -0.76 c^e_R\Bigr) \Lambda_{\rm TeV}^{-2},
\eeqa{numerics}
where $\Lambda_{\rm TeV}\equiv \Lambda/(1~{\rm TeV})$.
To estimate the sensitivity of Higgs factories to new physics, we consider two scenarios for the cross section measurement precision: a ``conservative" one, $\delta\sigma/\sigma=0.5$\%, and an ``optimistic" one, $\delta\sigma/\sigma=0.1$\%. (If statistical error dominates, the conservative scenario corresponds to an integrated luminosity $\int L dt\approx 180$~fb$^{-1}$, or about 3 years of running the ILC-250 at design luminosity. The optimistic scenario corresponds to $\int L dt\approx 4500$~fb$^{-1}$, which would probably require combining data from multiple detectors as envisioned, for example, in the TLEP proposal.) Table~\ref{tab:reach} and Fig.~\ref{fig:reach} show the exclusion and discovery reaches in a new physics scenario where a single dim.-6 operator dominates. For this estimate,
we only used the total cross section measurement, and assumed that it is in exact agreement with the reference value computed in the $(m_Z, G_F, \alpha)$ basis. Of course, this information can be augmented with angular distributions, asymmetries, {\rm etc.}, further improving the reach. We defer a consideration of such improvements to future work.

In addition to running at $\sqrt{s}=250$ GeV, the physics program of the Higgs factory may include running at higher energies as well; for example, in the case of the ILC, running scenarios including periods of running at 350 GeV and 500 GeV are being discussed. While a detailed analysis of the physics reach of such scenarios is beyond the scope of this paper, to facilitate future work we present the analogues of Eq.~\leqn{numerics} at these energies:
\beqa
\sqrt{s}=350~{\rm GeV}:~~~\frac{\Delta\sigma}{\sigma} &\approx& \Bigl( 0.36 c_{WW} + 0.01 c_{BB} +0.06 c_{WB} 
-0.06 c_H -0.04 c_T +2.01 c_L^{(3)\ell} \CR & & + 0.28 c_{LL}^{(3)\ell} + 1.73 c_L^\ell -1.48 c^e_R\Bigr) \Lambda_{\rm TeV}^{-2}, \CR
\sqrt{s}=500~{\rm GeV}:~~~\frac{\Delta\sigma}{\sigma} &\approx& \Bigl( 0.45 c_{WW} + 0.02 c_{BB} +0.08 c_{WB} 
-0.06 c_H -0.04 c_T +3.82 c_L^{(3)\ell} \CR & & + 0.28 c_{LL}^{(3)\ell} + 4.10 c_L^\ell -3.02 c^e_R\Bigr) \Lambda_{\rm TeV}^{-2}. 
\eeqa{numericsADD}
As expected, the contributions of most operators grow with energy, and better reach can be obtained if an equivalent sample of Higgs bosons is collected at higher energies. 

It is instructive to compare the sensitivity of the measurement discussed here with the current bounds on these operators, which come primarily from precision electroweak fits and the LHC measurements of the Higgs rates. The final two columns of Table~\ref{tab:reach} list the precision electroweak constraints, obtained from Ref.~\cite{Falkowski:2014tna}, and the bounds derived from the agreement of the CMS measurement of the effective $h\gamma\gamma$ vertex, $\kappa_\gamma$, with the SM~\cite{Khachatryan:2014jba}. (ATLAS constraints on this vertex are very similar~\cite{ATLAS-CONF-2014-009}.) For most operators, the sensitivity of the Higgs factory is well in excess of the current bounds, the only exceptions being $\CO_{BB}$ and, for the conservative luminosity assumptions, $\CO_{WB}$.

Another relevant question is how the Higgsstrahlung cross section will compare, in terms of new physics sensitivity, to various other observables that can be measured at the Higgs factory. The operators $\CO_{BB}$, $\CO_{WB}$ and $\CO_{WW}$ will be constrained by a precise measurement of $\kappa_\gamma$, to which they contribute as
\beq
\Delta\kappa_\gamma  = \frac{1}{2}\frac{\Delta \Gamma(h\to \gamma\gamma)}{\Gamma(h\to \gamma\gamma)} \approx -2.9 (c_{WW}+c_{BB} -c_{WB} ) \Lambda_{\rm TeV}^{-2}.
\eeq{hgammagamma_ops}
We estimate that a measurement of $\kappa_\gamma$ with 8\% precision, roughly corresponding to the ILC-500 projection of the Snowmass-2013 study~\cite{Dawson:2013bba}, would have a 95\%~c.l. exclusion reach of $\Lambda/\sqrt{c_i}\leq 4.3$ TeV for each of these operators. The same measurement with a 1.5\% precision, projected for TLEP in the same study, would increase the reach to about 10 TeV. This is comparable to the Higgsstrahlung sensitivities in the case of $\CO_{WW}$, and significantly exceeds the Higgsstrahlung reach for $\CO_{BB}$ and $\CO_{WB}$. However, we emphasize that the relative size of dimension-6 operators depends on the details of new physics at the scale $\Lambda$, and the Higgsstrahlung cross section gives access to several operators not accessible to other measurements.

\section{Third-Generation Squarks: The NLO Calculation}
\label{sec:NLO}

In this section, we analyze the corrections to $\HZ$ due to loops of third-generation squarks of supersymmetric (SUSY) models. There are two related reasons to focus on these particular contributions. First, third-generation squarks are required to be relatively light, below 1 TeV, to avoid the need for significant fine-tuning in the EWSB sector~\cite{Dimopoulos:1995mi,Cohen:1996vb,Perelstein:2007nx,Brust:2011tb,Papucci:2011wy}. Most other superpartners can be heavier without inducing fine-tuning. In fact, such a split spectrum, often referred to as ``Natural SUSY", is preferred in light of the strong LHC bounds on gluinos and squarks of the first two generations. Second, even if some other superpartners are below 1 TeV, the third-generation squark effects in $\HZ$ are enhanced due to the large value of the top Yukawa coupling.

We implemented the ``Natural SUSY" model in the {\sc FeynArts} package~\cite{Hahn:2000kx,Hahn:1998yk} by including the third generation left-handed doublet $\widetilde{Q}_3=(\widetilde{T}_L,\widetilde{B}_L )$ and right-handed singlet $\widetilde{T}_R$ fields within the SM model file. (The right-handed sbottom $\widetilde{B}_R$ does not have to be below 1 TeV to maintain naturalness, and we do not include it in the calculation.)
The three input parameters for the squark sector are the two soft masses $\widetilde{m}_L, \widetilde{m}_R$ and the $A_t$ trilinear soft coupling.  The D-term scalar potential is also included; however this does not introduce additional free parameters as the couplings are determined through the electroweak couplings. The Lagrangian is thus given by
\beqa
\mathcal{L} & = &  \mathcal{L}_{SM} + \mathcal{L}_{Kin,\tilde{t}} - \widetilde{m}^2_L |\widetilde{Q}_3|^2- \widetilde{m}^2_R |\widetilde{T}_R|^2 - A_t ( \widetilde{T}_R  H \cdot \widetilde{Q}_3 +h.c.) \\
& & -\lambda_t^2 |H|^2 (|\widetilde{Q}_3|^2+|\widetilde{T}_R|^2) - \frac{g'^2}{2} \left(\frac{2}{3} |\widetilde{T}_R|^22 -\frac{1}{6} |\widetilde{Q}_3|^2 -\frac{1}{2} |H|^2 \right)^2 \\
& & - \frac{g^2}{2} \sum_a \left(\widetilde{Q}^\dagger_3 \cdot \tau^a \cdot \widetilde{Q}_3 +H^\dagger \cdot \tau^a \cdot H \right)^2,
\eeqa{stoplag}
where $\tau^a = \sigma^a/2$ with $\sigma^a$ the usual Pauli matrices.  This Lagrangian can be obtained from the MSSM by decoupling all superpartners other than $\widetilde{Q}_3$ and $\widetilde{T}_R$, and taking the usual decoupling limit in the Higgs sector, $m_{H_d}^2\to \infty$ and $\tan\beta\to\infty$. Note, however, that  our implementation treats the Higgs mass $m_h$ as a free parameter, while in the MSSM it is not. This is motivated by the well-known tension between naturalness and the 125 GeV Higgs in the MSSM. The tension is reduced in extended models with additional tree-level contributions to $m_h$, such as the NMSSM or models with non-decoupling D-terms. Our implementation of $m_h$ ensures that our results are applicable in such models, in the limit when extra BSM states are decoupled. It should also be noted that due to the absence of A-terms mixing $\widetilde{B}_L$ with $\widetilde{B}_R$ the physical mass of  $\widetilde{B}_L$ is very close to $\widetilde{m}_L$, and hence whenever $\widetilde{m}_L \lesssim 120$ GeV we will assume that other bounds from direct searches are satisfied by additional mixings in the sbottom sector which raise the physical mass of both sbottoms.

In order to renormalize the theory for the calculation of virtual corrections, a minimum basis of three input parameters must be chosen, and then counterterms are defined for those parameters and the SM field strengths.  Due to the ease of implementation in the {\sc{FeynArts}}, {\sc{FormCalc}}, and {\sc{LoopTools}} suite of packages~\cite{Hahn:2000kx,Hahn:1998yk} we opt for the complete on-mass-shell renormalization scheme~\cite{Denner:1991kt,Aoki:1980ix,Aoki:1980hh,Aoki:1982ed} and hence choose electroweak inputs of $(M_Z, M_W, \alpha_{EM})$ following the prescription of~\cite{Denner:1991kt}.

The full set of counterterms includes the field strength counterterms, particle mass counterterms (including the Higgs mass counterterm, which we consider to be independent unlike in the MSSM), a counterterm for the EM coupling at low energies, and a counterterm for the Higgs vev.  All other counterterms are then defined through combinations of this set.  Due to their weak charges and couplings to the Higgs, the squarks enter into the counterterms for the weak sector.

\begin{figure}[t]
  \centering
 \includegraphics[height=0.4\textwidth]{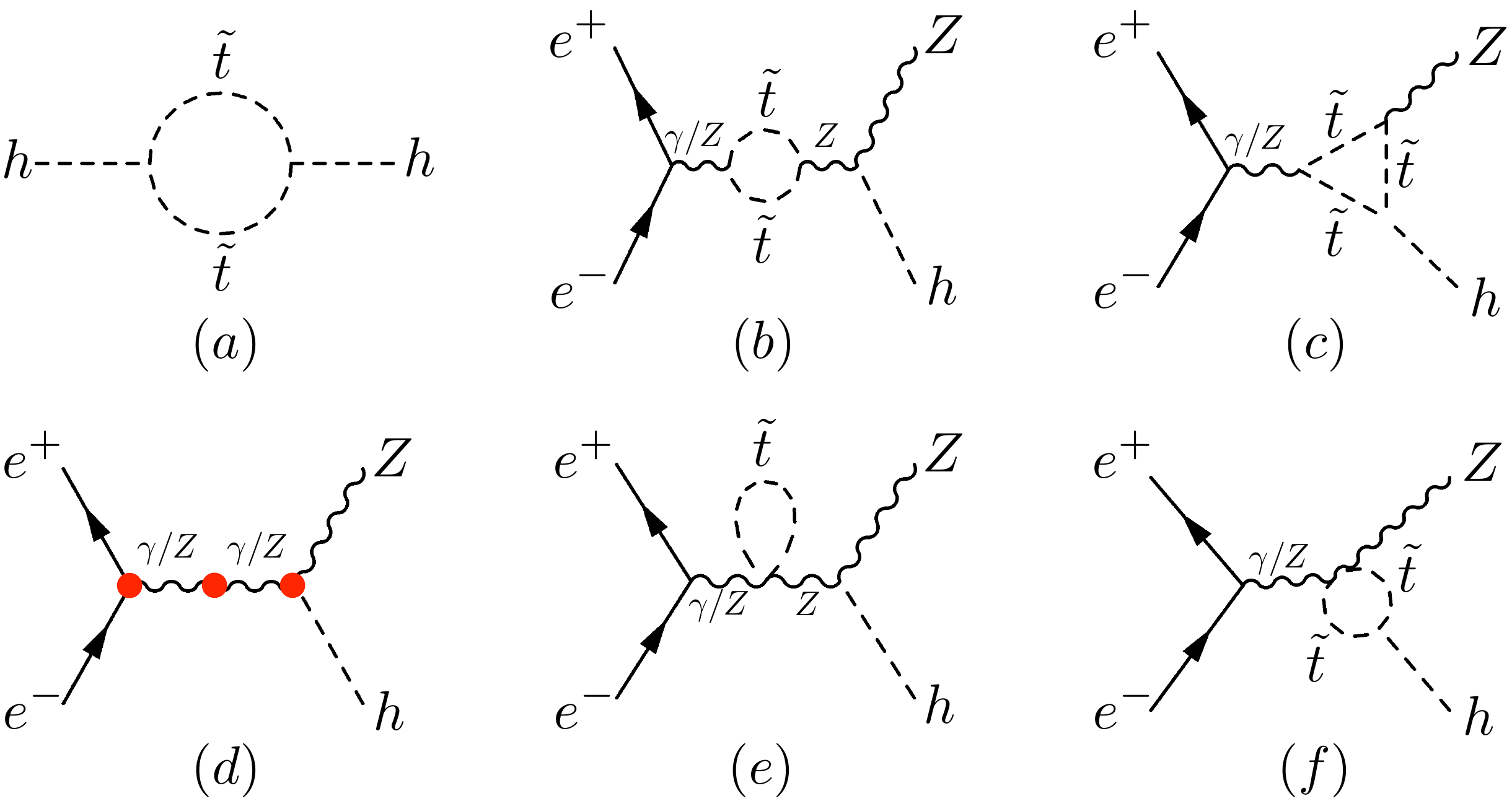}
  \caption{Feynman diagrams corresponding to $e^+e^- \to hZ$ at a lepton collider.  The Higgs wavefunction correction diagram discussed in~\cite{Craig:2013xia} is shown in (a), and all possible counterterm diagrams are shown in (d) with the understanding that in the calculation of one-loop counterterms only the stop and left-handed bottom squarks are included.  One-loop $Z/\gamma$ wavefunction correction diagrams (b,e) and vertex correction (c,f) diagrams are also shown.  Diagrams involving left-handed bottom squarks are not shown, but also contribute.}
  \label{fig:loopdiagrams}
\end{figure}

\begin{figure}[t]
  \centering
 \includegraphics[height=0.4\textwidth]{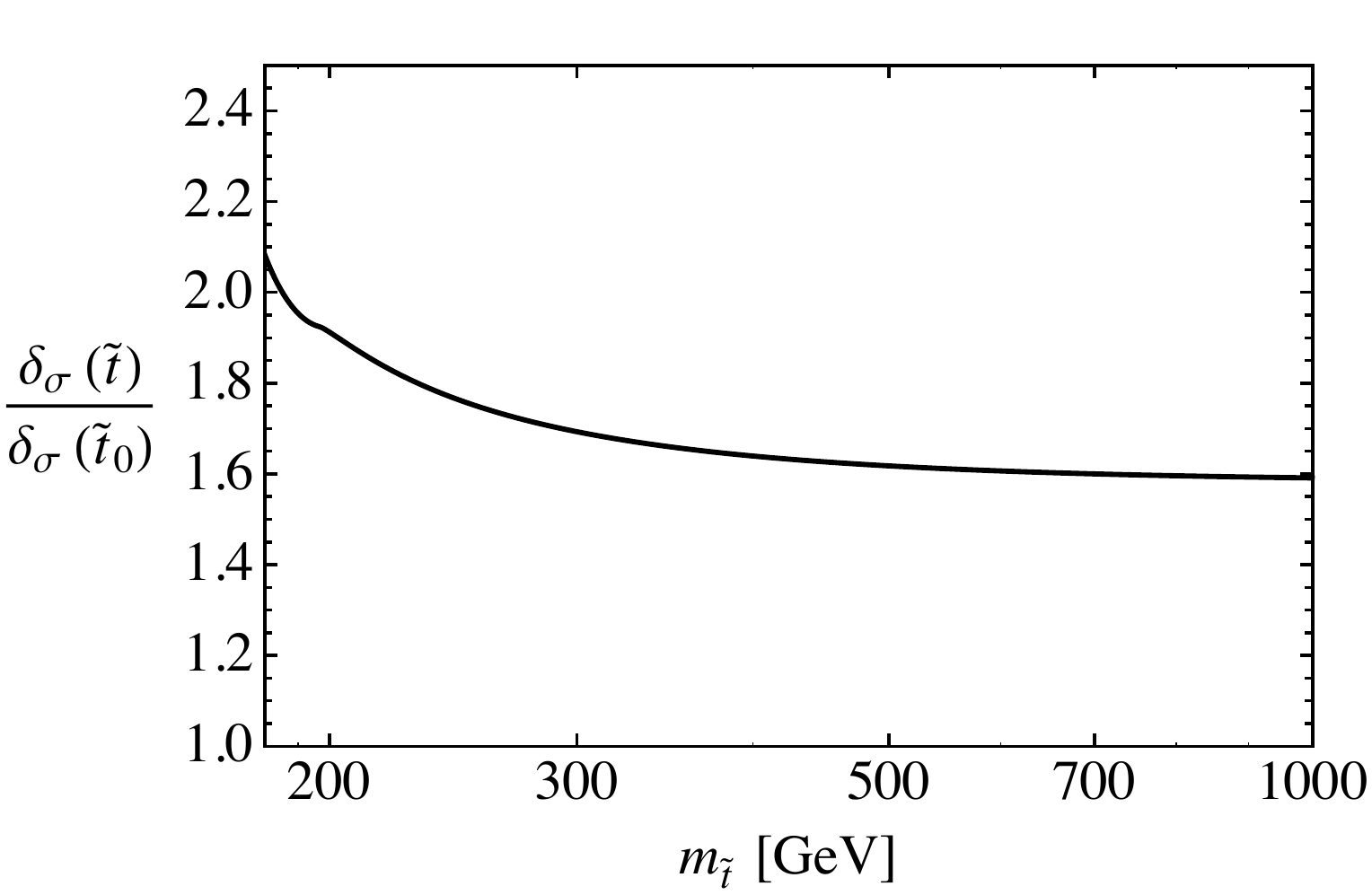}
  \caption{The ratio of the full NLO correction from stops/sbottoms ($\tilde{t}$) relative to the correction from gauge-singlet scalar top partners ($\tilde{t}_0$).
  }
  \label{fig:comparisonnaive}
\end{figure}

Some of the NLO diagrams contributing to $\HZ$ are shown in Fig.~\ref{fig:loopdiagrams}. It has been analytically checked that the full NLO correction is finite and gauge invariant. As a further check, setting stop couplings to gauge bosons to zero in our NLO calculation reproduces the results of~\cite{Craig:2013xia}, where gauge-singlet scalars $\tilde{t}_0$ were considered.\footnote{For proper comparison, $A_t$ must be set to 0 in the stop case, since it has no counterpart in the case of $\tilde{t}_0$; and the $\tilde{t}_0$ correction must be rescaled by 6 to account for the number of degrees of freedom in the two stop fields.} In the case of $\tilde{t}_0$, the effect arises entirely from the quartic coupling $\lambda_t^2 |H|^2 |\tilde{t}_0|^2$, which induces an irreducible physical correction to the Higgs wavefunction renormalization, Fig.~\ref{fig:loopdiagrams} (a), and the corresponding counterterm. Naively, one might expect this contribution to dominate the NLO correction for stops as well, since it is enhanced relative to other diagrams by the ratio of the top Yukawa to gauge couplings. However, this is not the case, as can be seen in Fig.~\ref{fig:comparisonnaive} which compares the full NLO correction from stops to the correction from $\tilde{t}_0$. The Higgs wavefunction renormalization accounts for $50-70$\% of the full NLO result, depending on the stop mass. We conclude that the electroweak interactions of stops do play an important role in the NLO contribution.

\subsection{Comparison between EFT and Full-NLO Predictions}
\label{sec:compare}

As the EFT and NLO calculation methods must agree in the heavy-squark limit, the combination of both methods allows for a non-trivial cross check of the results.  There is additional interplay between the two as the NLO calculation allows the regions of validity of the EFT calculation to be clearly determined, and on the other hand the EFT calculation allows for some physical insights into the results of the NLO calculation.  The comparison is also interesting as although the differences between EFT results and loop functions in LO processes such as $gg\to h$ or $h\to \gamma \gamma$ have been thoroughly studied, the interplay between EFT and NLO results, where the full systematics of renormalization are at play, is much less studied and makes the comparison interesting in a formal sense.

\begin{figure}[t]
  \centering
 \includegraphics[height=0.32\textwidth]{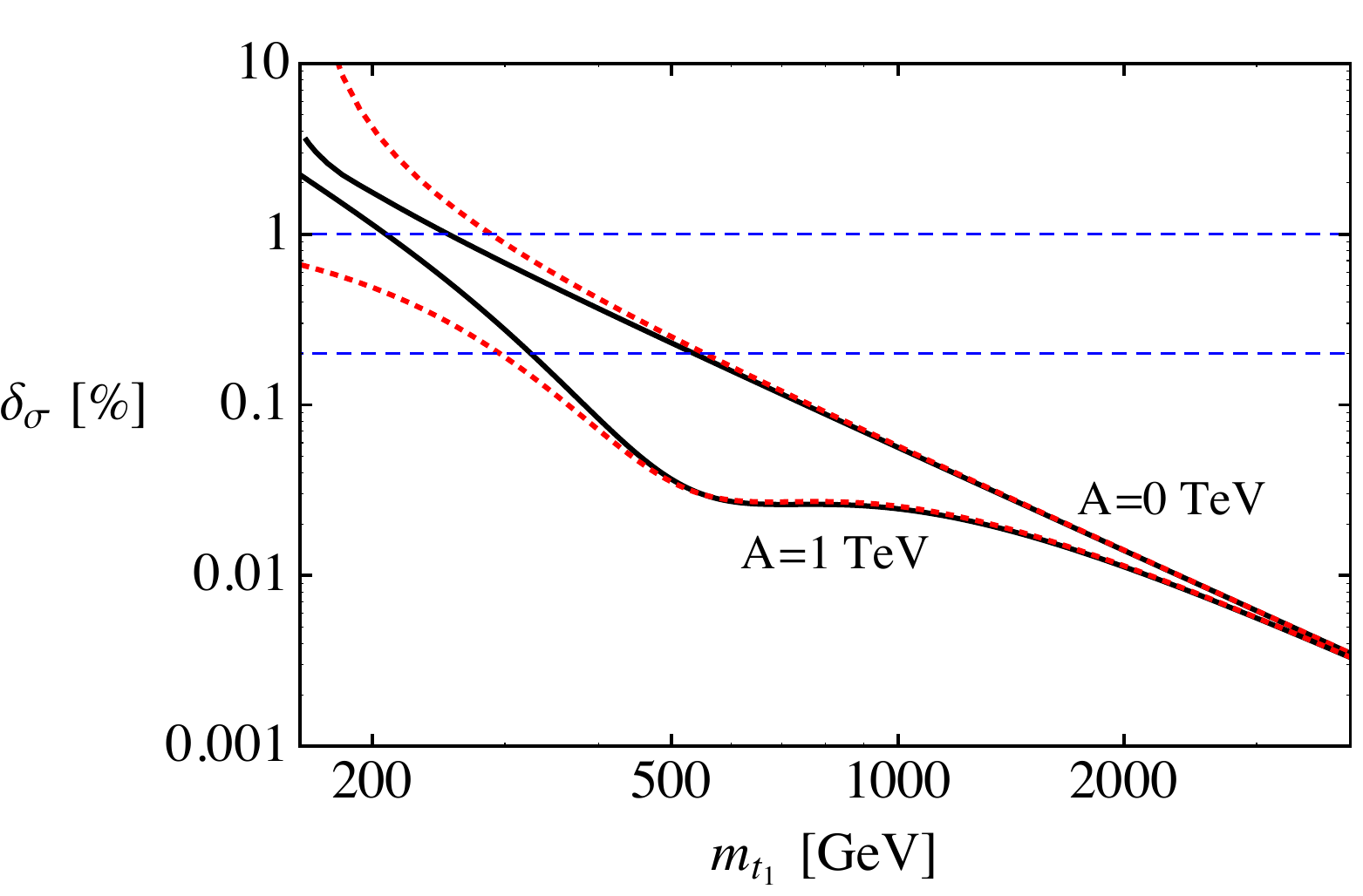} \, \, \includegraphics[height=0.32\textwidth]{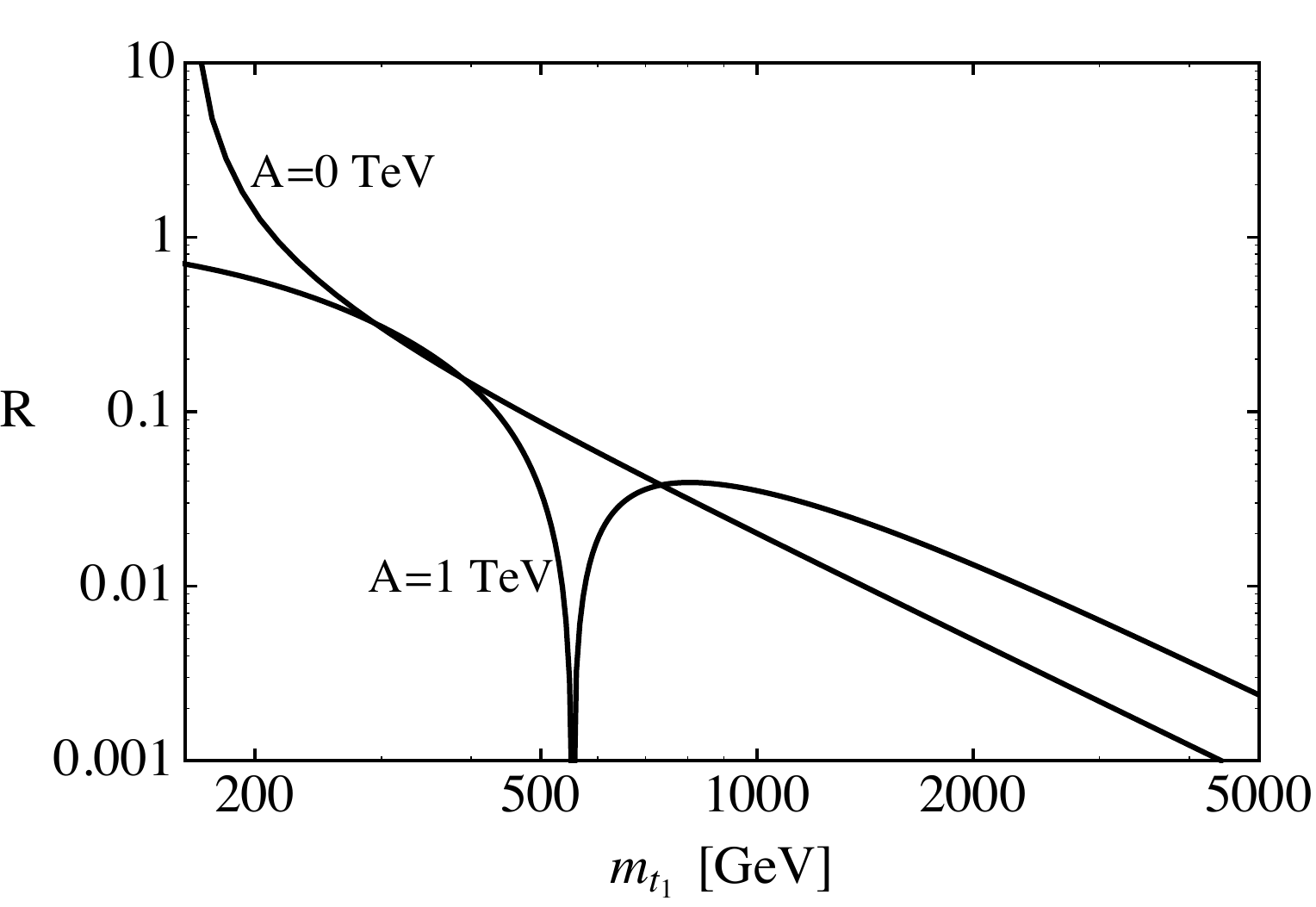}
  \caption{Left: Fractional corrections to the Higgsstrahlung cross section as a function of the physical mass $\widetilde{m}_1$ of the lightest stop squark, for equal soft masses $\widetilde{m}_L=\widetilde{m}_R$ and two values of the A-term.  NLO results are shown in solid black and EFT results in dashed red.  For comparison the conservative and optimistic estimates of the $2\sigma$ reach of a Higgs factory are shown in dashed blue.  Right:  The ratio of EFT to NLO results $R=\delta_{\sigma}^{EFT}/\delta_{\sigma}^{NLO}-1$ for the same parameters.
  }
  \label{fig:comparison}
\end{figure}

\begin{table}[t]
\begin{center}
\begin{tabular}{|r|c|}
\hline
$c_{WW}$  &  $\hat{c}_{WW}$\\
$c_{BB}$  &  $\hat{c}_{BB} $\\
$c_{WB}$   &$\hat{c}_{WB} $  \\
$c_H$ & $\hat{c}_H - \hat{c}_R + \frac{3 }{4} g^2 \hat{c}_{2W}-\frac{3 }{2} g^2 \hat{c}_{W}$\\
$c_T$ & $\hat{c}_T + \frac{1}{4} g'^2 \hat{c}_{2B}-\frac{1}{2} g'^2 \hat{c}_{B}$\\
$c_L^{(3)\ell}$ & $- \frac{1}{4} g^2 \hat{c}_{2W}+ \frac{1}{4} g^2 \hat{c}_{W}$ \\
$c_{LL}^{(3)\ell}$ & $- \frac{1}{8} g^2 \hat{c}_{2W}$\\
$c_L^\ell$ & $\frac{1}{4} g'^2 \hat{c}_{2B}-\frac{1}{4} g'^2 \hat{c}_{B}$\\
$c_R^e$ & $\frac{1}{2} g'^2 \hat{c}_{2B}-\frac{1}{2} g'^2 \hat{c}_{B}$ \\
\hline
\end{tabular}
\end{center}
\caption{A dictionary to translate the Wilson coefficients in Table~II of Ref.~\cite{Henning:2014gca}, denoted here by $\hat{c}_i$, into coefficients $c_i$ of the operators in our basis.}
\label{tab:dictionary}
\end{table}%

To perform this comparison, we utilize the results of Ref.~\cite{Henning:2014gca}, where the Wilson coefficients of all relevant dim.-6 operators induced by third-generation squark loops have been calculated at the one-loop order. Equal soft masses, $\widetilde{m}_L=\widetilde{m}_R\equiv \widetilde{m}_S$, have been assumed; we will only consider this limiting case in this subsection. The basis of dim.-6 operators used in Ref.~\cite{Henning:2014gca} is slightly different from the one we use, Table~\ref{tab:hit}. Using equations of motion, we obtain a dictionary to translate the results of~\cite{Henning:2014gca} into our basis, shown in Table~\ref{tab:dictionary}. The EFT prediction for the $\HZ$ cross section is then obtained by inputting the $c_i$ coefficients into the formulas of Sec.~\ref{sec:eft}. It is important to use the same input basis in the EFT and NLO calculations; in our case, it is the basis $(m_Z, m_W, \alpha)$.

The EFT and NLO calculations are at the same order (one-loop) in the usual perturbation theory in gauge and Yukawa couplings. The EFT result is in addition leading-order in the expansion in inverse powers of $\widetilde{m}_S$, while the NLO result is exact in $\widetilde{m}_S$. Thus we expect the discrepancy between the two to scale approximately as the ratio of $\widetilde{m}_S$ and the other mass scales in the calculation, such as $m_Z, m_H, v,$ and $\sqrt{s}$.  As the CM energy $\sqrt{s}$ is the largest of the relevant energy scales, we estimate that the difference between NLO and EFT calculations should be $\mathcal{O}(s/\widetilde{m}^2_s)$, and the two should converge rapidly in the heavy squark limit.

In Fig.~\ref{fig:comparison} we show both results for two choices of the soft trilinear coupling $A_t=0,~1$ TeV.  The two results indeed converge rapidly, becoming virtually indistinguishable when the lightest stop mass exceeds $\widetilde{m}_1 \gtrsim 500$ GeV.  The difference between the two approaches scales as $\propto 1/\widetilde{m}^2_S$ as expected, confirming that the other details of the calculation, in particular the Wilson coefficients of the effective operators, are in excellent agreement.

In Fig.~\ref{fig:comparison} it is also clear that in the light-stop region accessible to a potential future Higgs factory, the EFT calculation may significantly over- or underestimate corrections to Higgstrahlung.  This behavior is particularly notable in the case with vanishing A-term. The EFT expansion in the inverse powers of $\widetilde{m}_S$ breaks down when $\widetilde{m}_S\lsim \sqrt{s}$. In the case of $A_t=0$, physical stop masses $\widetilde{m}_{1,2} \sim m_t$ actually imply very small soft masses $\widetilde{m}_S \sim 0$, since $\widetilde{m}_{1,2} = m_t$ in the limit of exact SUSY. For non-zero A-terms this situation may be avoided, as small physical masses are possible with relatively large (weak-scale) soft masses by tuning the A-term contribution against the soft mass contribution. In this case, the EFT expansion remains valid to accuracies $\sim \mathcal{O}(10\text{'s}\%)$.  In summary, in the parameter space which may be accessible to a future Higgs factory the NLO calculation is desirable for an accurate prediction; however, as long as the soft masses are not vanishing, the EFT calculation serves as a useful approximation.

\subsection{Higgs Factory Reach for Stops}
\label{sec:StopReach}

\begin{figure}[t]
  \centering
 \includegraphics[height=0.375\textwidth]{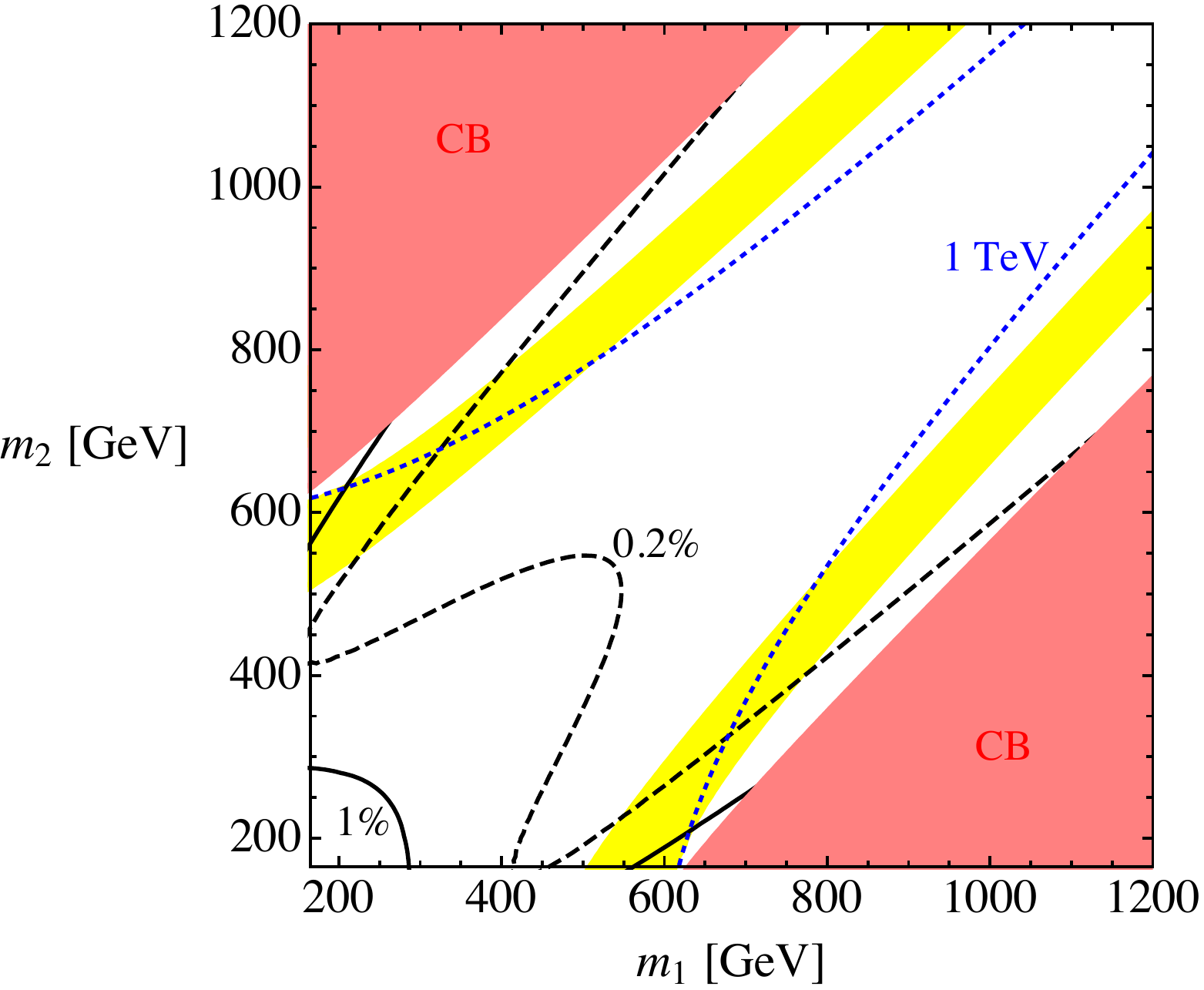} \, \, \includegraphics[height=0.375\textwidth]{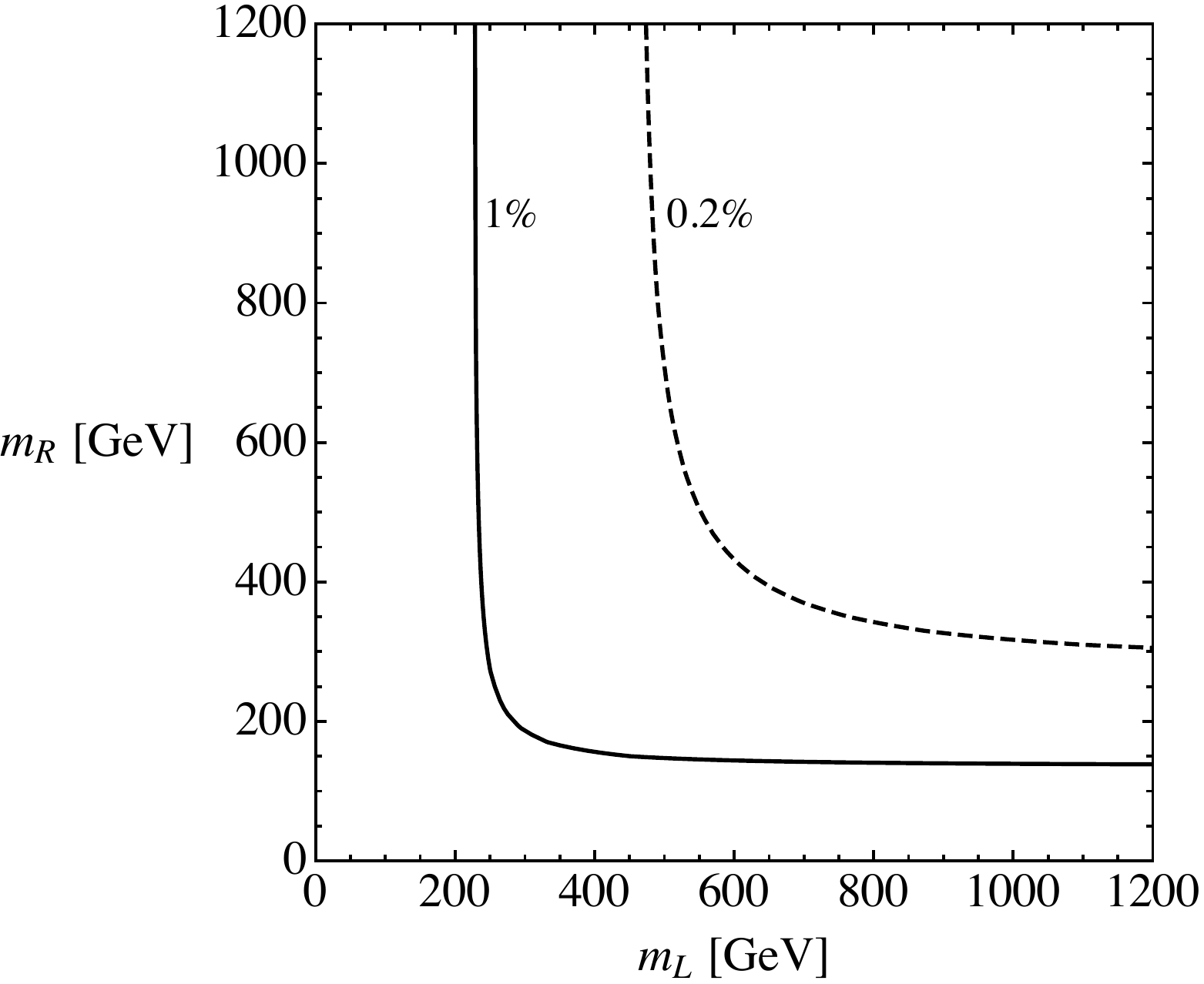}
  \caption{Conservative (solid black) and optimistic (dashed black) estimates of the $2\sigma$ reach of a Higgs factory through precision Higgstrahlung measurement.  Left: Equal soft masses $\widetilde{m}_L=\widetilde{m}_R$, with physical masses plotted on each axes.  Contours of constant A-term are also shown in dotted blue, and regions with color-breaking vacua have been shaded out in pink. In the yellow bands, the observed Higgs mass
 is realized within the MSSM. Right: NLO results for unequal soft masses and vanishing A-term, with the physical left and right-handed stop masses shown on the axes.  It should be noted that in all regions of parameter space the corrections to the Higgsstrahlung cross section are negative.
  }
  \label{fig:stopreach}
\end{figure}

As there are three input parameters, $\widetilde{m}_L, \widetilde{m}_R$ and $A_t$, a full characterization of the parameter space would require a three dimensional scan.  In Fig.~\ref{fig:stopreach} two well-motivated `slices' of this parameter space are presented.  On the left-hand plot the two soft masses are set equal $\widetilde{m}_L=\widetilde{m}_R=\widetilde{m}_S$, and the trilinear A-term is varied. On the right-hand plot the A-term is set to zero and both soft masses are varied. In both cases, $2\sigma$ Higgs factory reach contours, in the conservative and optimistic scenarios, are projected onto the plane of physical stop masses.

We first consider the left-hand plot.  In the region with a small A-term ($m_1 \sim m_2$) the conservative (optimistic) experimental reach extends to about $250~(500)$ GeV. As the A-term is increased, the sensitivity decreases. The reason is suppression of the Higgs coupling to the lightest stop squark, due to a cancellation between the coupling from the trilinear term and the coupling from the quartic term.  This leads to a ``blind spot" around the line $m_1 = m_2 \pm \sqrt{2} m_t$. However, as the A-term is increased further, the cancellation no longer persists and the sensitivity increases again. The maximum size of the A-term is limited by the requirement that the theory must possess no color-breaking vacua. This constraint implies that the maximal possible size of the effect in Higgsstrahlung cross section is about 1\%. For illustration, we also show the region of the parameter space where the MSSM Higgs mass prediction is $m_h = 125 \pm 2$ GeV. (This prediction is subject to significant theoretical uncertainty, since it is obtained using the two-loop leading-log approximation of Ref.~\cite{Carena:1995bx} for stop loops, and does not include contributions from other MSSM particles.) An observable shift in the Higgsstrahlung cross section is predicted in parts of that region.

On the right-hand plot the difference between the left- and right-handed squark contributions is illustrated.  It is clear that the corrections due to left-handed stops exceed those from right-handed stops.  This is perhaps not surprising as the right-handed stops only couple to the Higgs and hypercharge. In theories with small A-terms the estimated experimental reach is $\sim225~(475)$ GeV for left-handed stops alone, and $\sim170~(250)$ GeV for right-handed stops.

An extensive program of direct searches for stops is currently underway at the LHC. If R-parity is conserved and the stop-LSP mass splitting is large, $m_{\tilde{t}}-m_{\rm LSP} \gg m_t$, the current bounds on the stop mass are already about $700-750$ GeV, well in excess of the Higgs factory reach we found. However, direct searches depend crucially on the spectrum of the SUSY particles, and on the stop decay channels. For example, in the R-parity conserving case, for stop and LSP masses in the regions $m_{\tilde{t}}-m_{\rm LSP} \approx m_t$ or $m_{\tilde{t}}-m_{\rm LSP} \approx m_W$, stops below 200 GeV are allowed by direct searches. Additional constraints on light stops in this region have recently been placed by high-precision measurements of the $t \bar t$ cross section \cite{Aad:2014kva} and $t \bar t$ spin correlations \cite{ATLASspin}. The $t \bar t$ cross section measurement excludes stops between $m_t < m_{\tilde t} < 177$ GeV assuming the decay $\tilde t_1 \to t \tilde \chi_1^0$ proceeds predominantly to right-handed top quarks. Although this does provide a weak limit for light stops, it may be entirely eroded by mixed branching ratios, three-body decays, or changes in the LSP identity. The $t \bar t$ spin correlation measurement excludes stops between $m_t < m_{\tilde t} < 191$ GeV, likewise assuming the decay $\tilde t_1 \to t \tilde \chi_1^0$ proceeds to predominantly right-handed top quarks with $m_{\tilde \chi_1^0} = 1$ GeV. Dependence of the limit on $m_{\tilde \chi_1^0}$ is relatively weak, but as in the case of the cross section limit it may be substantially eroded by mixed branching ratios, three-body decays, or changes in the LSP identity. Also, removing the assumption of R-parity conservation drastically weakens the bounds. For example, if the stops decay to two jets via the RPV $UDD$ operator, all stop events result in purely hadronic final states buried under the large QCD background. Currently there is no LHC bound on this scenario \cite{Bai:2013xla}. In both cases, the difficulty faced by direct searches is not statistics -- in fact the LHC Run-1 would already have produced a large sample of stops in these scenarios -- but rather the difficulty of separating signal from background. This indicates that these scenarios will remain challenging for the LHC in Run-2 and beyond, and may well still be unconstrained at the time the Higgs factory becomes operational.

\begin{figure}[t]
  \centering
 \includegraphics[height=0.375\textwidth]{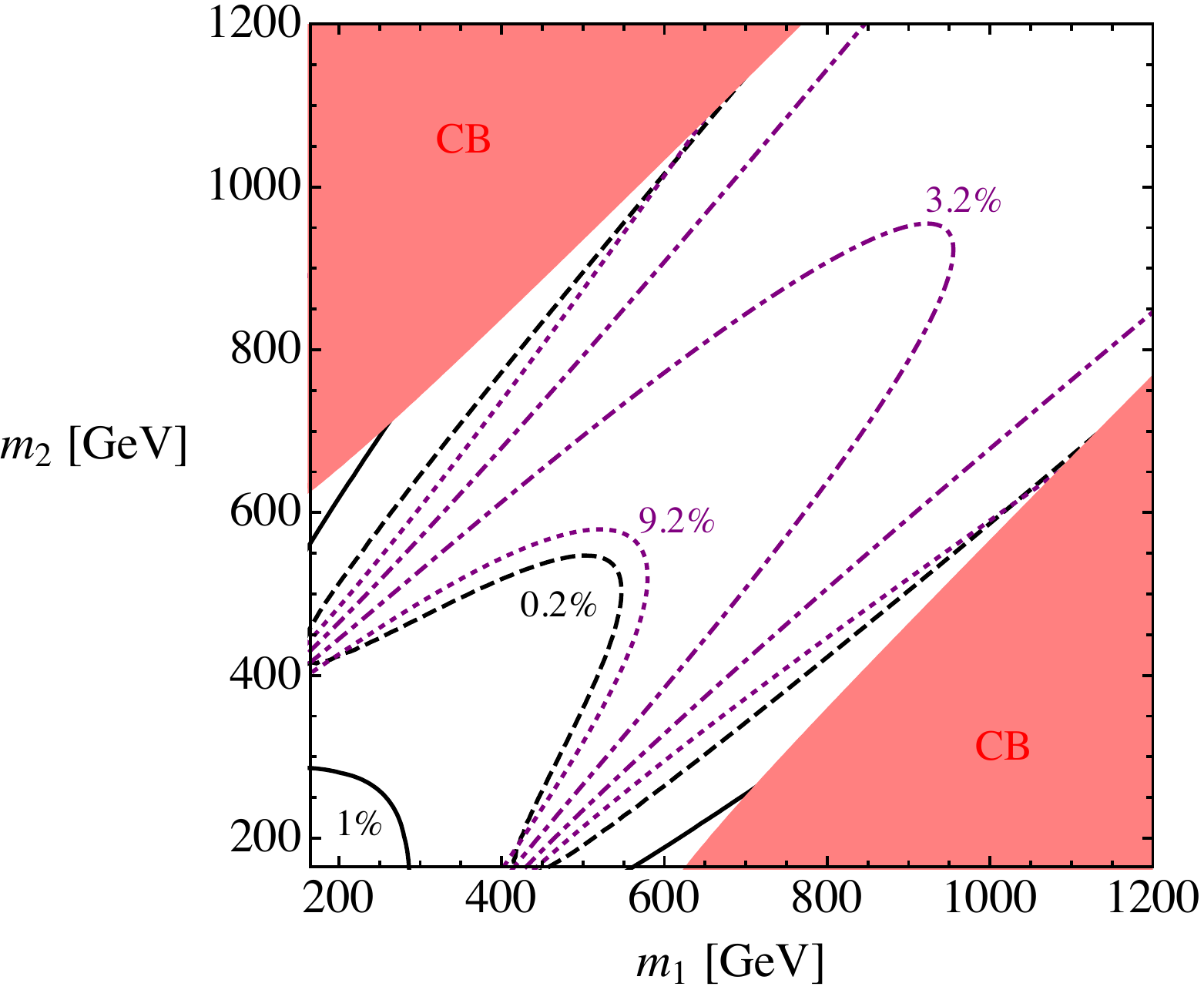} \, \, \includegraphics[height=0.375\textwidth]{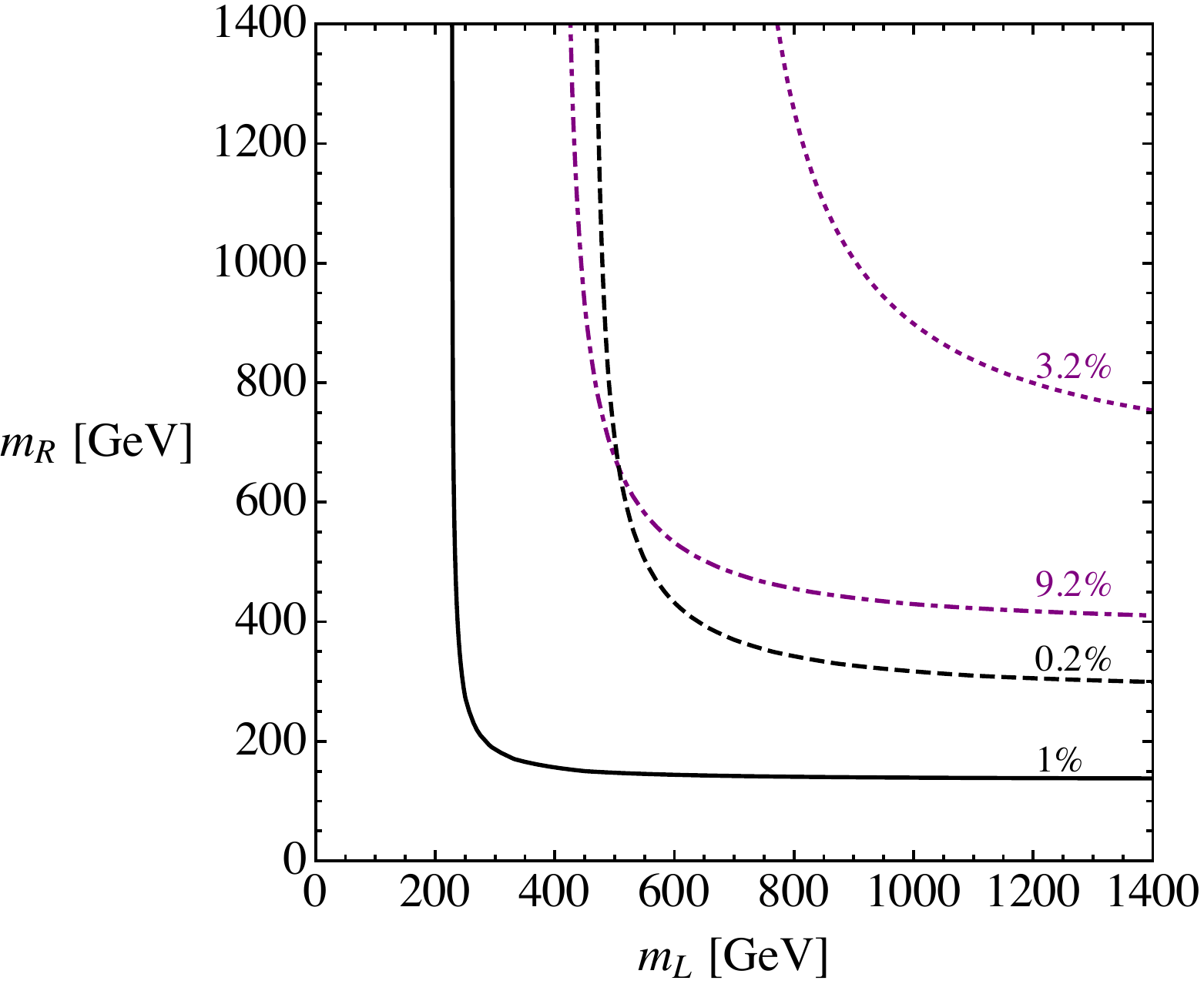}
  \caption{As in Fig.~\ref{fig:stopreach} with the addition of projected conservative (dotted) and optimistic (dotdashed) sensitivity from $h\to gg$ measurement shown in purple.  Absolute deviations are shown but it should be kept in mind that the $h\to gg$ corrections are positive for small A-terms (central region of left-hand plot and all of right hand plot) and negative for large A-terms (large mass splittings).}
  \label{fig:gluecomp}
\end{figure}

In addition to Higgsstrahlung measurement, the Higgs factory can perform other indirect searches for stop squarks.  The leading sensitivity would be due to modifications of the $h\to gg$ decays arising from loops of heavy stops. To compare the sensitivities, we again consider a ``conservative" and an ``optimistic" scenario, assuming a 4.6\% and 1.6\% precision in the measurement of $\Gamma (h \to gg)$. (The two numbers correspond to the estimates of the Snowmass report~\cite{Dawson:2013bba} for the ``ILC-500" and ``TLEP-350" scenarios, respectively.)
In Fig.~\ref{fig:gluecomp} we compare the reach of the Higgs factory for stops via $h\to gg$ and Higgsstrahlung measurements. It is clear that $h\to gg$ has the higher sensitivity throughout the parameter space. Note that the ``blind spot" is common for both measurements, since it is due to the suppression of the $h\tilde{t}_1\tilde{t}_1$ coupling which affects equally both channels. Thus, the Higgsstrahlung measurement unfortunately cannot be used to eliminate or shrink this gap in the coverage of $hgg$. However, both measurements probe interesting regions of parameter space, and they are complementary in that while $hgg$ is only sensitive to the color quantum numbers of stops, the Higgsstrahlung is sensitive to their electroweak quantum numbers.

\begin{figure}[t]
  \centering
 \includegraphics[height=0.375\textwidth]{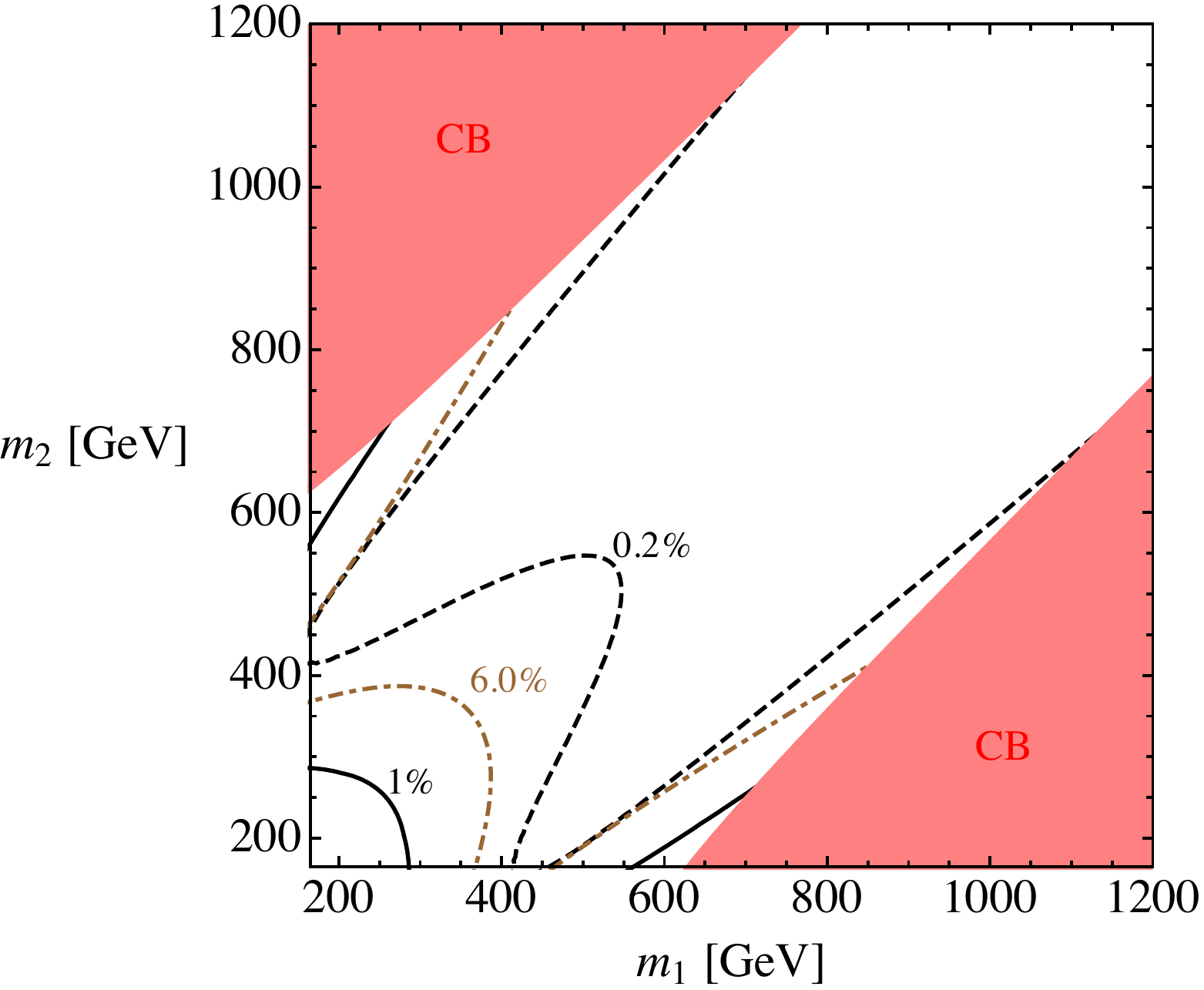} \, \, \includegraphics[height=0.375\textwidth]{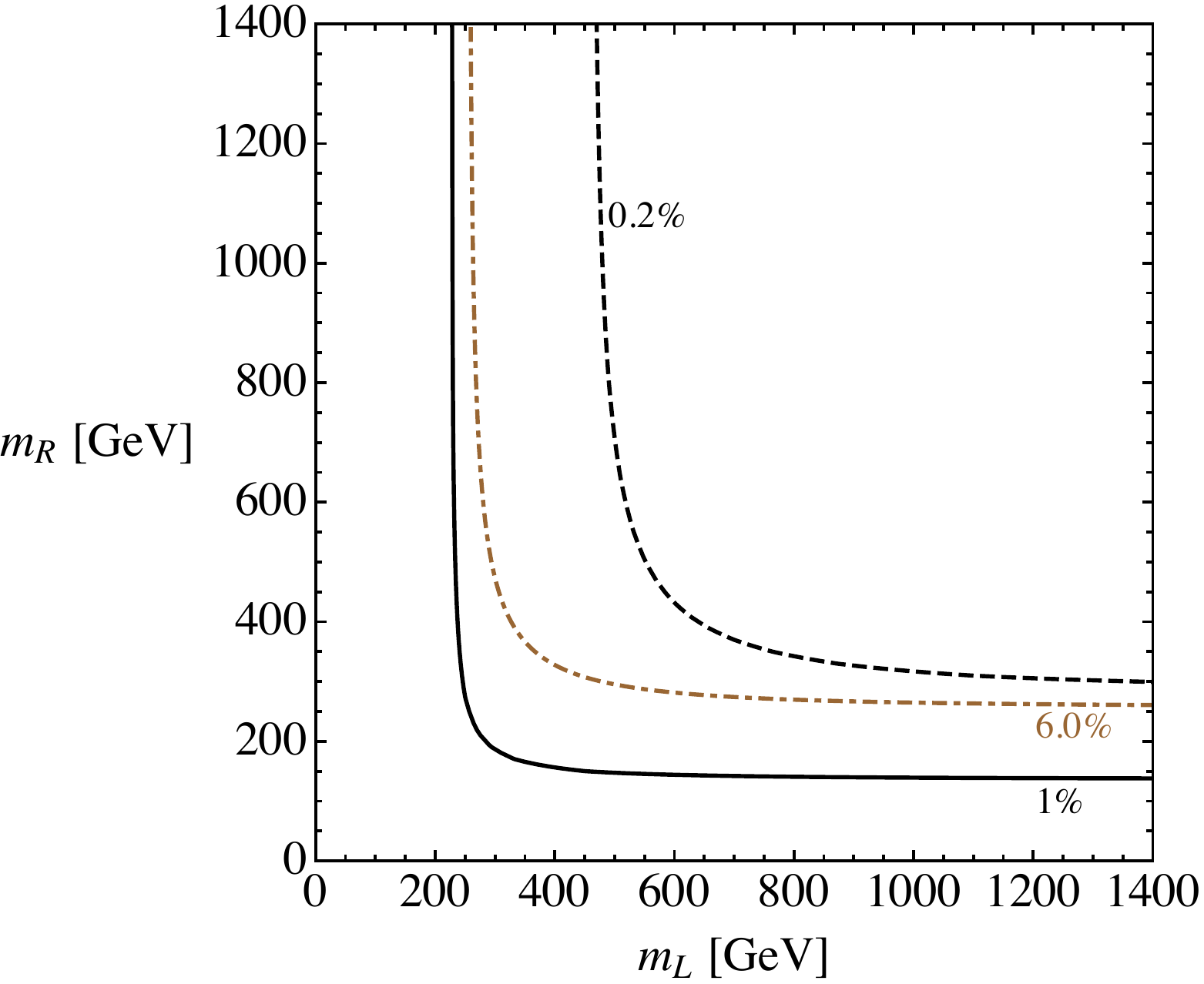}
  \caption{As in Fig.~\ref{fig:stopreach} with the addition of projected optimistic (dotdashed) sensitivity from $h\to \gamma\gamma$ measurement shown in brown.  Absolute deviations are shown but it should be kept in mind that the $h\to \gamma\gamma$ corrections are negative for small A-terms (central region of left-hand plot and all of right hand plot) and positive for large A-terms (large mass splittings).}
  \label{fig:gammacomp}
\end{figure}

Sensitivity to the electroweak quantum numbers of top partners is particularly useful in more exotic scenarios such as folded supersymmetry~\cite{Burdman:2006tz}, in which scalar top partners (the $F$-stops) are charged under Standard Model electroweak interactions but neutral under QCD. The $F$-stop corrections to $\HZ$ are identical to those of stops, but there is no corresponding modification to $h \to gg$. Rather, the competing indirect probe would be modifications to $h \to \gamma \gamma$ decays arising from loops of $F$-stops, which is less strongly constrained than $h \to gg$ at a Higgs factory. We consider an optimistic baseline scenario, corresponding to a $3.0$\% precision in the measurement of $\Gamma (h \to \gamma\gamma)$.  (This corresponds to the estimate of the Snowmass report~\cite{Dawson:2013bba} for the ``TLEP-350" scenario.)  Fig.~\ref{fig:gammacomp} compares the reach of the Higgs factory for $F$-stops via $h\to \gamma\gamma$ and Higgsstrahlung measurements. As with $h \to gg$ and Higgsstrahlung, $h \to \gamma\gamma$ exhibits the common ``blind spot'' due to suppression of the $h \tilde{t}_1 \tilde{t}_1$ coupling. However, in contrast to $h \to gg$, $h \to \gamma \gamma$ has lower sensitivity than Higgsstrahlung across the parameter space. This is due to both the smaller numerical coefficient for $h\gamma\gamma$ corrections relative to $hgg$ corrections at a given point in parameter space, and the weaker fractional sensitivity of the Higgs factory to $h\gamma\gamma$. Thus in exotic scenarios such as folded supersymmetry, Higgsstrahlung provides the most sensitive indirect tool to search for top partners at the Higgs factory.

\section{Conclusions}
\label{sec:conc}

A very precise measurement of the Higgsstrahlung cross section can be performed at a future Higgs factory. In this paper, we considered the potential of this measurement to search for new physics. First, we computed the shift in the cross section due to a complete set of effective dim.-6 operators that can contribute. Second, we performed a complete NLO calculation of the cross section shift due to third-generation squarks in supersymmetric models. We established that the two calculations agree in the limit of large stop masses, providing a highly non-trivial check on both sides. We also discussed the physics reach of this measurement. In the case of dim-6 operators induced at {\it tree-level}, we find that the typical scale of new physics that can be probed is of order a few TeV, with the precise number depending on the operator, as well as the assumed measurement precision (see Table~\ref{tab:reach} and Fig~\ref{fig:reach}). In the case of stops, we found that masses up to about 500 GeV can be probed under the best-case scenario, see Fig.~\ref{fig:stopreach}. The weaker sensitivity for stops is due to the fact that they only affect the cross section at the one-loop order. Using the projections from the Snowmass study~\cite{Dawson:2013bba}, we find that the measurement of $\Gamma(h\to gg)$ will provide a more sensitive indirect probe of stops than the Higgsstrahlung cross section measurement at a Higgs factory. This remains true throughout the model parameter space: in particular, in the ``blind spots" where the stop contribution to $\Gamma(h\to gg)$ vanishes due to an accidental cancellation, their contribution to Higgsstrahlung is also highly suppressed and cannot be used to probe this region. On the other hand, in models such as ``Folded SUSY", where the top partners have electroweak quantum numbers as stops but are not colored, a Higgsstrahlung cross section measurement does provide the best sensitivity, beating $\Gamma(h\to \gamma\gamma)$.

\section*{Acknowledgements}
The authors are grateful for conversations with Christoph Englert, Andrey Katz, Hong-Yu Ren, Francesco Riva,  Jesse Thaler, and Carlos Wagner.  M.M. acknowledges support from a Simons Postdoctoral Fellowship and a CERN COFUND Fellowship. M.M. and N.C. acknowledge support from the Aspen Center for Physics and NSF grant 1066293 where this work was partially completed.  MF and MP are supported by the U.S. National Science Foundation through grant PHY-1316222 and CAREER grant PHY-0844667.

\bibliography{lit}
\bibliographystyle{jhep}

\end{document}